\documentstyle[preprint,aps,eqsecnum]{revtex}

\def\beq#1{\begin{equation} \label{#1}}
\def\eeq{\end{equation}}
\newcommand{\bea}{\begin{eqnarray}}
\newcommand{\eea}{\end{eqnarray}}
\def\bra#1{\left\langle #1\right\vert}
\def\ket#1{\left\vert #1\right\rangle}
\def\epsp{\epsilon^{\prime}}                    
\def\NPB{{ Nucl. Phys.} B}
\def\PLB{{ Phys. Lett.} B}
\def\PRL{ Phys. Rev. Lett.}
\def\PRD{{ Phys. Rev.} D}
\def\AJP{{\em Am. J. Phys.}}
\begin{document}
{
\tighten

\title {New method for studying neutrino mixing and mass differences} 
\author{Harry J. Lipkin\,\thanks{Supported in part by 
U.S.
Department of Energy, Office of Nuclear Physics, under contract 
number
DE-AC02-06CH11357.}} 
\address{ \vbox{\vskip 0.truecm}
  Department of Particle Physics
  Weizmann Institute of Science, Rehovot 76100, Israel \\
\vbox{\vskip 0.truecm}
School of Physics and Astronomy,
Raymond and Beverly Sackler Faculty of Exact Sciences,
Tel Aviv University, Tel Aviv, Israel  \\
\vbox{\vskip 0.truecm}
Physics Division, Argonne National Laboratory,
Argonne, IL 60439-4815, USA\\
~\\harry.lipkin@weizmann.ac.il
\\~\\
}

\maketitle

\begin{abstract} 

A toy model shows how neutrino masses and mixing can be investigated by
studying the behavior of a radioactive ion which decays by K-capture BEFORE and
DURING its weak decay by K-capture. A new oscillation phenomenon providing
information about neutrino mixing is obtained by following the ion before and
during the decay.  This normally neglected process is shown to be consistent
with quantum mechanics and causality. Measuring the oscillation without
detecting the neutrino avoids losses in conventional experiments due to the low
neutrino absorption cross section. The normally unobservable long wave lengths
are made observable by having the radioactive source move a long distance
circulating around in a storage ring.  The initial ion wave packet has a
momentum spread required by Heisenberg and contains pairs of components with
different momenta and energies. These can produce neutrino amplitudes in two
mass eigenstates with different momenta which mix to produce a single
$\nu_e$ state.  In this typical quantum mechanics ``two-slit" or
``which path"  experiment a transition between the same initial and final
states can go via two paths in energy-momentum space with a phase difference
producing interference and oscillations.

\end{abstract}

} 


\def\beq#1{\begin{equation} \label{#1}}
\def\eeq{\end{equation}}
\def\bra#1{\left\langle #1\right\vert}
\def\ket#1{\left\vert #1\right\rangle}
\def\epsp{\epsilon^{\prime}}
\def\NPB{{ Nucl. Phys.} B}
\def\PLB{{ Phys. Lett.} B}
\def\PRL{ Phys. Rev. Lett.}
\def\PRD{{ Phys. Rev.} D}
\section{Introduction}

\subsection{The two principal difficulties of neutrino experiments}

A recent experiment\cite{gsi} describes an oscillation observed in 
the decay of a radioactive ion before and during the emission of an unobserved
neutrino. This phenomenon offers a new and very interesting method for
determining neutrino masses and mixing angles\cite{gsikienle,gsifaber}. 

\begin{enumerate}
\item Ordinary neutrino oscillation experiments are difficult because
\begin{itemize}    
 \item The neutrino absorption cross section is tiny. The number of neutrino
 events actually used in ordinary experiments is many orders of magnitude smaller than the 
 number events creating the neutrinos. 
 \item The oscillation wave lengths are so large that it is difficult to 
actually follow even one oscillation period in any experiment.
\end{itemize}  
\item 
This experiment opens up a new line for
dealing with these difficulties
\begin{itemize}  
\item The oscillation is measured without detecting the neutrino.
Detection of every neutrino creation event
avoids the losses from the low neutrino absorption cross section.
   \item The long wave length problem is solved by having the
radioactive source move a long distance circulating around in a storage
ring.  The data if
correct show many oscillations in the same experiment.

\end{itemize}
\end{enumerate}

 This paper considers the  basic quantum mechanics of the first difficulty and
shows in a toy model that it is possible in principle to observe and measure
neutrino oscillations by looking only at the radioactive source. 
 The second difficulty warrants further investigation. 

The theoretical analysis in this paper was motivated by discussions with  Paul
Kienle at a very early stage of the experiment in trying to understand whether
the effect was real or just an experimental error. 

\subsection{The K-capture experiment}
The original version of this paper was written on the basis of private
information before the release of ref.\cite{gsi} and considered the decay of
a nucleus by emitting an electron into the atomic K-shell. 

A similar anaysis can be applied to a K-capture experiment in which a 
radioactive ion in an atom or ion decays by capturing an electron  from the
K-shell or other atomic shell and emits a monoenergetic  neutrino. Here there
are a number of initial states having different  degrees of ionization.
Interference can only occur between initial states  having the same degree  of
ionization. Otherwise the analysis is the same  as for the K-emission
experiment. However, we note that the paths in space  for a neutral atom are
not easily influenced by external electromagnetic  fields that can otherwise
influence the path or orbit in space taken by  the ion in the initial state.

The emitted electron-neutrino $\nu_e$ is now known to be a linear combination
of several neutrino mass eigenstates.  If the initial state has a definite
momentum and energy, the conservation of energy and momentum determines the
energy and momentum of the neutrino and therefore its mass. This is then a
``missing mass" experiment in which the mass of the neutrino is determined
without the observation of the neutrino. Interference between amplitudes from
different neutrino mass states cannot be observed in such a missing-mass
experiment. 

At first it seems rather  peculiar that neutrino oscillations can be observed
in the state of a radioactive ion before its decay into an unobserved neutrino.
One wonders about causality and how the initial ion can know how it will decay.
But much discussion and thought revealed that the essential quantum mechanics 
is a ``two-slit" or ``which-path" experiment\cite{leofest} in momentum space
which preserves causality  because no information about the final state is
available to the initial ion.

It is not a missing mass experiment because
the initial ion wave function is a wave packet containing a combination of
momenta which prevent it from being used in a missing mass experiment. Its
well-defined relative phases are determined by its localization in
space. These relative phases change with time in accordance with the relative
energy differences in the packet. 

The weak decay transition then produces a final neutrino via any of its
momentum eigenstates. Since the same final state can be produced by any of the
momentum components in the initial wave function, the path in energy-momentum
space between the initial and final states is not known and the corresponding
amplitudes can be coherent and interfere.    

The relative phases in the initial wave function are independent of the final
state, which is created only at the decay point. Thus there is no violation of
causality. No information about the final state exists before the decay.
Although time-dependent perturbation theory might suggest that a decay
amplitude can be present before the decay, the continued observation of the
initial ion before the decay rules out any influence of any final state
amplitude on the decay process. It is like the  ``Schroedinger cat" experiment
in which the door is always open so that there is a continuous measurement of
whether the cat is still alive.

\subsection{The quantum mechanics of realistic experiments}

In any realistic experiment the
Heisenberg uncertainty principle prevents the momentum of the initial state
from being known with sufficient precision to determine the neutrino mass.  
The GSI experiment\cite{gsi} observed periods of modulation of the order of 7
seconds with ions traveling at 71\% of the velocity of light. The ions thus
travel a distance of $0.71\cdot 3 \cdot 10^5  \cdot 7 \approx 10^6$ kilometers
in a single period of oscillation. The uncertainty  in knowing the position  of
the experiment within the laboratory is tiny incomparison with this enormous
oscillation wave length.  Heisenberg then tells us that the momentum
uncertainty required to produce thes oscillations is equally tiny in comparison
to the momentum fluctuations required by confining the experiment within the
laboratory. 
\beq{heisen}
\frac{\delta x}{\lambda_{osc}} \approx \frac{\delta p_{osc}}{\delta p_{loc}} 
\ll 1  
\eeq
where $\delta x$ denotes the uncertainty in the position of the experiment in
the laboratory, $\lambda_{osc}$ denotes the oscillation wave length,  
$\delta p_{osc}$ denotes the momentum difference required for these oscillations
and $\delta p_{loc}$ the momentum difference in the initial state required by
its localization in the laboratory. Thus this is not a missing-mass experiment. 

The momentum difference between the different neutrino mass states
is thus  much smaller than the momentum uncertainly required by Heisenberg from
knowing that the experiment takes place within the laboratory. The initial
state is a wave packet in momentum space containing the different momenta
required to produce decays to neutrino mass eigenstates with different masses.
The transition to the final $\nu_e$  state can therefore go via
different neutrino mass eigenstates with no record of which mass eigenstate
produced the final $\nu_e$. The contributions via different neutrino mass 
eigenstates define different paths in momentum space which are not observed in
the experiment. The contributions to the final state amplitude via these
different paths are therefore coherent and interference between them can be
observed producing oscillations.

\subsection{Coherence and decoherence}

Understanding coherence and decoherence is crucial here. 
All the relevant physics is in the initial state of the ion.
The amplitude at the decay point is the  coherent sum of the amplitudes from
all allowed paths in energy-momentum space.  But coherence between amplitudes
is not introduced by simple ignorance of which path was taken\cite{Kurt}.
Coherence results only from an uncertainty required by quantum mechanics.
Most nuclear and
particle theorists are unfamiliar with investigations on coherence and
decoherence in condensed matter and mesoscopic physics\cite{ADY,pwhichfin}. 
In neutrino oscillation experiments the answer is clear. The
neutrino is detected, time is not measured and the detector has a momentum
uncertainty. The relevant neutrino states are those of the same energy and
different momentum. These are the only pairs of states where coherence can
be preserved.

This experiment\cite{gsi} is completely different.  Time is measured and a time
dependence is the crucial new ingredient in the experiment. However, the
preparation of the initial state is complicated and considerable thought has to
be devoted to some questions raised in the paper\cite{gsi}. A radioactive
ion is trapped in a storage ring with a circumference of 108.3 m and a
revolution frequency of about 2 MHz.  Time oscillations in the radioactive
decay by emission of an unobserved neutrino are observed with a period of about
10 seconds. How could coherence be preserved over the time span of some ten
seconds? What is the effect of the continuous monitoring of the state of the
ion?

Condensed matter theorists have examined coherence and decoherence in many 
contexts.  There are also the concepts of ``preselection" and ``postselection" 
introduced and extensively explored by Yakir Aharonov and collaborators.

The purpose of this paper is to describe the basic quantum mechanics in a toy
model where the initial radioactive ion is moves in a straight line and oscillations are
observed as a function of time. This toy model is highly unrealistic.  A
precise analysis of the real experiment enabling the determination of neutrino
mass differences from the observed oscillation periods in a storage ring is
left for further investigation. It is much more complicated than this toy
model.    

The toy model provides some insight into how the coherence is preserved and 
the effect of continuous monitoring. In the model the initial state is a  wave
packet of plane waves which is moving in space and time. But plane waves have a
constant amplitude over all space and have an equal probability of being behind
the moon as in the laboratory where the state is created. The relative phases
of the individual plane wave components must be adjusted so that there is no
probability of finding the particle outside the laboratory at this time. The
center of mass momentum of the packet can change appreciably in the
preparation and cooling of the state. But these interactions cannot suddenly
create a probability that the ion has jumped to behind the moon. The limits on
the size of the wave packet in space preserve the relative phase of neighboring
components with the tiny difference in momenta relevant to the observed
oscillations. 

\section{The quantum mechanics of oscillation experiments}

\subsection{No coherence in a missing mass experiment}

A radioactive ion that decays by K-capture emits a neutrino which is a 
linear combination 
of neutrino mass eigenstates. If the energy and momentum of the initial 
ion and also the recoil momentum of the final ion are known
the energy and momentum of the emitted neutrino are 
determined and therefore its mass. This would suggest that there can be 
no coherence nor oscillations between the neutrino states.

To see that this argument misses the exciting observable physics 
from the beta-decay experiment we examine the analogous case of the 
emission of an electron and a neutrino in the decay of a pion at 
rest.\cite{leofest,pnow98}  

When a pion decays at rest $\pi \rightarrow e \nu$ the energies 
$E_e,E_\nu$
and momenta  $\vec p_e, \vec p_\nu$ of the electron and neutrino  
can all
be known. This is then just a ``Missing Mass" experiment. The neutrino mass $M_\nu$ is
uniquely determined by $M_{\nu}^2 = (M_{\pi} - E_e)^2 - p_e^2$. So 
how
can there be coherence and interference between states of different mass? We
are guided to the resolution  of this paradox by experience in condensed matter
physics discussing which amplitudes are coherent in quantum
mechanics\cite{ADY,Leo,Kayser}.

\subsection{Why it is not a missing mass experiment}

The original Lederman-Schwartz-Steinberger experiment found that the neutrinos
emitted in a $\pi-\mu$ decay produced only muons and no electrons. Experiments
now show that at least two neutrino mass eigenstates are emitted in $\pi-\mu$
decay and that at least one of them can produce an  electron in a neutrino
detector. The experimentally observed absence of electrons can be explained
only if the electron amplitudes received at the detector from different
neutrino  mass eigenstates are coherent and exactly cancel. This implies  that
sufficient information was not available to determine the  neutrino mass from
energy and momentum conservation. A missing mass experiment was not  performed.

Coherence or interference between different neutrino mass eigenstates 
cannot be observed in a ``missing mass" experiment where the mass of an 
unobserved neutrino is uniquely determined by other measurements and 
momentum and energy conservation.

The resolution of these  contradictions is just simple quantum mechanics.  In
any experiment which can detect neutrino oscillations, the position of  the
source must be known with an error much smaller than the wave length  of the
oscillation to be observed. The quantum mechanical uncertainty  principle
therefore forces coherence between neutrino mass eigenstates  having the same
energy and different momenta. Stodolsky's theorem\cite{Leo}  states that in
an experiment which does not explicitly measure time the quantum mechanical
density matrix for the system is diagonal in energy and there can be
interference between states of different energy and no explicit time dependence
in a correct theoretical description. The location in space already says it
all.

\section{The K-capture experiment in a toy model}

\subsection{The basic theory} 

The experiment describes the decay of a radioactive ion into another
radioactive ion by K-capture and the emission of a neutrino. Since there are
two different neutrino mass states, the decay has two neighboring channels for
decay. The standard theoretical description here is a linear combination of two
Breit-Wigner amplitudes. In a conventional experiment in which the momentum and
energy of the initial ion is known, the momentum and energy of the recoil ion
can be measured, the neutrino mass is determined and there can be no
oscillations. 

We have seen that in pion decay the localization of the pion in space produces
an uncertainty in momentum that prevents a determination of the neutrino mass.
The absence of a time measurement requires\cite{Leo} that only states of the
same energy can be coherent.

In the K-capture experiment\cite{gsi}   
the localization of the experiment in the laboratory also produces
quantum momentum fluctuations. But many other factors
are very different and much more complicated:
\begin{enumerate} 
\item Time is measured and interference between states of
different energies can be observed. 
\item The initial state is a single-particle state with a well defined mass.
There is therefore an uncertainty in energy required by the energy-momentum 
relation for a particle with a definite mass. The broadening of the mass value
by the small decay width is neglected here.
\item Oscillations are observed in the initial state, not the final state, as a
result of the time dependence of relative phases in the initial wave function. 
The experiment is a ``which-path experiment" because which particular momentum
eigenstate in the initial wave function produced the neutrino is not known. 
\item The final neutrino is not observed. It is known to have been created as
an electron neutrino which is a well-defined linear combination of two or three 
mass eigenstates. Further measurements on the final neutrino cannot affect the
oscillations.
\item The initial state is moving in a circular orbit in a magnetic field.
The kinematics may not be simple because the vector potential of the magnetic
field complicates the definition of momentum and momentum conservation and can
also introduce Aharonov-Bohm phases normally neglected. 
\item Lorentz transformations to bring the momentum of the initial ion to an
approximate rest system are not simple because the direction of the Lorentz
tranformation must change with the motion of the ion around the ring.
\end{enumerate}

In this paper the crucial question of which amplitudes are coherent is
considered in the framework 
of a toy model. 

\subsection{The K-capture experiment as a ``Two-slit" experiment}

This model
demonstrates that oscillations can occur and shows how an 
analysis of a realistic  experiment requires a detailed investigation of what
can be measured, what  can not and where coherence can occur. The relation
between the observed interference pattern and the neutrino masses is determined
by which terms are coherent.  The kinematics of the toy model is examined by
noting the following conditions for coherence. 

\begin{enumerate}

\item The initial state is a wave packet where Heisenberg prevents the
measurement of its energy and momentum.

\item The final neutrino is a coherent combination of two neutrino mass states.

\item The toy model used here assumes that momentum and energy of the final 
recoil ion is observable, whether it
is measured or not. Therefore states with different recoil momentum and energy
cannot be coherent.  

\item An open remaining question is the possible energy
non-conservation in times short compared to the time necessary to resolve the
two neutrino energy states. In this case the final state has an entangled 
wave function with two components having different recoil energies and different
neutrino masses. This possibility is outside the framework of our present toy
model but must be seriously considered for realistic cases.

\end{enumerate}

The crucial ingredient is the inability to know the momentum 
of the initial state because we know where it is and Heisenberg tells us 
that this requires an uncertainty in its momentum. 

The initial and final states of this experiment are well defined. The 
initial state is a radioactive ion in a wave packet which is confined
to definite region of configuration space and therefore has a momentum 
spread. The final state is a recoiling atom and the $\nu_e$ 
linear combination of the neutrino mass eigenstates produced 
when an electron disappears in the weak interaction. The location in space
of the initial and final states is well defined within normal experimental 
errors. 
We now have a direct analog to the two-slit or which-path experiment. 
Here the transition can go via any of the neutrino mass eigenstates. 
These define different paths between the initial and final states. 

The initial state is a wave packet with neither a sharp momentum nor a 
sharp energy. The waves on the two paths thus overlap in momentum and 
energy. 
Coherence can be observed only if we do not know which paths contributed 
to the transition. 

This requires that  the two paths have the have the the same recoil momentum
and energy, which are  observable in the final state.  The neutrino mass
eigenstates have  therefore different momenta which Heisenberg tells us must
be  unobservable. Since momentum is conserved in the transition, the  different
paths require different momenta for the radioactive ion in the  initial state.
The momentum spread in the initial wave functis  sufficient in practical
experiments to suppress all information on  ``which path" in momentum space was
taken in the transition.  

     The final $\nu_e$ state is a linear combination of mass 
eigenstates with different energies and different momenta and a well defined 
phase. During the time interval between the creation of the initial state and
its decay the relative phases between the energy eigenstate components of  the
initial wave function change linearly with the time. Thus the  probability that
the decay will take place to the final $\nu_e$ oscillates with
time. The period of the oscillation depends upon  the momentum and energy
differences which in turn depend upon the mass  differences between the mass
eigenstates.

      The experimental observation of these oscillations provides a new
experimental method to obtain information about the neutrino mass differences
and the mixing angles of the neutrino mass matrix.

\subsection{The kinematics of the transition}

Both energy and momenta are conserved for each component of the wave  packet
which has a momentum $\vec P$ and energy $E$ in the initial state.  The final
state has a recoil ion with momentum denoted by $\vec P_R$ and  energy $E_R$
and a neutrino with mass $m$, energy $E_\nu$ and momentum  $\vec p_\nu$.  The
energy release in the transition at rest, $M-M_R$ is denoted by $Q$. The
conservation laws then require

\beq{epcons} E_R= E - E_\nu;  ~ ~ \vec P_R = \vec P - \vec p_\nu ; ~ ~ 
M^2 + m^2 - M_R^2 = Q \cdot [2M-Q] + m^2 =2EE_\nu - 2\vec P\cdot\vec p_\nu 
\eeq 

We neglect transverse momenta and set 
$\vec P\cdot\vec p_\nu \approx P p_\nu$ where $P$ and $p_\nu$ denote the
components of the momenta in the direction of the incident beam. 
Then 
\beq{pnu}
 Q \cdot [2M-Q] + m^2 = 2E(E_\nu -p_\nu) + 2(E-P)p_\nu 
= \frac{2 E m^2}{E_\nu+p_\nu}+ 2(E-P)p_\nu  
\eeq 
This can be rearranged for further application to give
\beq{pnuenu}
 \frac {p_\nu}{P} - \frac {E_\nu}{E}=
\frac {p_\nu (E - P) + (p_\nu - E_\nu) P}{PE}= 
\frac {Q \cdot [2M-Q] + m^2}{2PE}-
\frac{m^2}{E_\nu+p_\nu}\cdot\left[\frac{1}{P}+ \frac{1}{E}\right]\ll 1 
\eeq

We are interested in the changes in the kinematic variables
$\delta p_\nu$, $\delta P$, $\delta E_\nu$ and $\delta E$ produced by a small
change $\Delta (m^2)$ in the squared neutrino mass: 

\beq{delm2}
\frac{\Delta (m^2)}{2} =E(\delta E_\nu) + (\delta E)E_\nu
 - P (\delta p_\nu) - (\delta P) p_\nu 
 \eeq

We assume that the interference occurs between two neutrino states with the same
energy and different momenta and that there is no change in the recoil momentum. 

\beq{sameng}
\delta E_\nu =0; ~ ~ ~ \delta p_\nu = \delta P = \frac{E}{P}\cdot \delta E
; ~ ~ ~ \frac{\delta p_\nu}{\delta E} = \frac{\delta P}{\delta E } =\frac{E}{P}
\eeq 

Combining eqs.(\ref{pnuenu}),
(\ref{delm2}) and
(\ref{sameng}) then gives

\beq{delm2pot}
\frac{\Delta (m^2)}{2\delta E} 
=E_\nu - P \frac{\delta p_\nu}{\delta E} - p_\nu\frac{\delta P}{\delta E } =  
 - E 
 \cdot \left[1 + 
\left \{\frac {p_\nu}{P} - \frac {E_\nu}{E}\right\}\right] \approx  - E
 \eeq

The phase difference at a time t between states produced by the neutrino
mass difference on the motion of the initial ion in the laboratory frame with 
velocity $V=(P/E)$ is
\beq{delphipotalt}
\delta \phi \approx -\delta E\cdot t                                                     
\approx  - \frac{\Delta (m^2)}{2E}\cdot t 
= - \frac{\Delta (m^2)}{2\gamma M}\cdot t 
\eeq                                                                                                                                                                
where $\gamma$ denotes the Lorentz factor $E/M$.
The period for $\delta \phi = - 2\pi$ is
\beq{periodalt}
\delta t \approx  
\frac{4 \pi \gamma M}{\Delta (m^2)}
\eeq

In ref.\cite{gsikienle} the period of modulation obtained was given as
\beq{periodirk}
\delta t_{IRK} = \frac{8 \pi \gamma M_R}{\Delta (m^2)} 
\eeq

The ratio of these two values is

\beq{periodrath}
\frac{\delta t}{\delta t_{IRK}} \approx 
\frac{M}{2M_R}\approx \frac{1}{2}
\eeq

Neither of these two values should be taken seriously because of essential
features neglected in the simple models.
Furthermore, the initial ion is not free but is constrained by electromagnetic
forces confining it to an orbit in a  storage ring. These forces introduce
potential energies which may be important in the kinematics. They
also complicate any Lorentz transformation from a ``rest system" to the
laboratory system. Simple attempts to include such effects have so far been
unsatisfactory and are not considered here. A full calculation may be necessary
including these confining forces.

One interesting feature of these two estimates is that they are within an order
of magnitude of the result obtained from neutrino experiments. This requires the
period to have a scale defined by the ratio of the mass of the ion to the
difference between the squared masses of the two neutrino states. Other
derivations and attempts to ``correct" unjustified approximations destroy this
scaling by introducing the very different energy scale of the neutrino energy.

The question arises of a possible additional phase proportional to the 
distance along the path. 
If we consider the motion along a straight line path, 
The total relative phase between waves differing by a momentum 
$\delta P$
for traversing a distance $X$ with velocity $V$ is 
\beq{delphipotalt2}
\delta \phi_{str} \approx \delta P\cdot X - \delta E\cdot t
\approx
\left[\frac {E}{P} \cdot V - 1\right] 
\delta E\cdot t =0                                                     
\eeq                                                                                                                                                                

But if the motion is in a circular orbit in a magnetic field with a  frequency
independent of the momentum there is no additional phase  accumulated in the
motion around the ring and eqs(\ref{periodalt}) and (\ref{periodrath}) apply.
The question of the phase difference between states of different momenta along
the two slightly different paths in a storage ring can thus play a crucial role
here and depends upon the experimental  conditions.  
 
\section{Conclusions}

A new oscillation  phenomenon providing information about neutrino mixing is 
obtained by following the initial radioactive ion  before and during the decay.
The difficulties introduced in conventional neutrino experiments by the tiny
neutrino absorption cross sections and the very long oscillaton wave lengths  
are avoided here. Measuring the decay time enables every neutrino event to be
observed and counted without the necessity of observing the neutrino via the
tiny absorption cross section. The confinement of the initial ion in a storage
ring enables long wave lengths to be measured within the laboratory.

Coherence between amplitudes produced by the weak decay of a radioactive  ion
by the emission of  neutrinos with different masses has been shown to follow
from the localization of the initial radioactive ion within a space
interval much  smaller than the oscillation wave length.  This coherence is
observable in following the motion of the initial radioactive ion from its
entry into the apparatus to its decay. The amplitude  for  production of a
$\nu_e$ from several mass eigenstates depends upon the relative
phases of the contributions from components in the inital wave function having 
different energies and momenta. These relative phases increase linearly with
time and produce oscillations.

Observing the period of these oscillations 
gives information about the neutrino mass differences and the mixing angles of
the neutrino mass matrix. Reliable detailed values for the relation between the
observed oscillation period and neutrino mass differences are not obtained in
the crude models so far considered.   At this point the fact that the values
obtained (\ref{periodalt}) and (\ref{periodirk}) are within an order of
magnitude of consistency with values obtained\cite{gsikienle} from neutrino
oscillation experiments is encouraging.   

\section{Acknowledgement}

It is a pleasure to thank Paul Kienle for calling my attention to this problem
at the Yukawa Institute for Theoretical Physics at Kyoto  University, where
this work was initiated during the YKIS2006 on ``New  Frontiers on QCD".
Discussions on possible experiments with Fritz Bosch, Walter Henning, Yuri
Litvinov and Andrei Ivanov are also gratefully acknowledged along with a
critical review of the present manuscript. The author also acknowledges further
discussions on neutrino oscillations as ``which path" experiments with  Eyal
Buks, Maury Goodman,  Yuval Grossman, Moty Heiblum, Yoseph Imry, Boris Kayser,
Lev Okun, Gilad Perez, David Sprinzak, Ady Stern, Leo  Stodolsky and Lincoln
Wolfenstein,

%
\catcode`\@=11 
\def\references{ 
\ifpreprintsty \vskip 10ex
%
\hbox to\hsize{\hss \large \refname \hss }\else 
\vskip 24pt \hrule width\hsize \relax \vskip 1.6cm \fi \list 
{\@biblabel {\arabic {enumiv}}}
{\labelwidth \WidestRefLabelThusFar \labelsep 4pt \leftmargin \labelwidth 
\advance \leftmargin \labelsep \ifdim \baselinestretch pt>1 pt 
\parsep 4pt\relax \else \parsep 0pt\relax \fi \itemsep \parsep \usecounter 
{enumiv}\let \p@enumiv \@empty \def \theenumiv {\arabic {enumiv}}}
\let \newblock \relax \sloppy
 \clubpenalty 4000\widowpenalty 4000 \sfcode `\.=1000\relax \ifpreprintsty 
\else \small \fi}
\catcode`\@=12 
{\tighten

} 


\begin{thebibliography}{99}
\bibitem{gsi} Yu.A. Litvinov, H. Bosch et al
nucl-ex0801/2079
\bibitem{gsikienle} A.N.Ivanov, R.Reda and P.Kienle
nucl-th0801/2121
\bibitem{gsifaber} Manfred Faber
nucl-th0801/3262
\bibitem{leofest} {Harry J. Lipkin.
hep-ph/0505141, 
Phys.Lett. B642 (2006) 366}

\bibitem{Kurt}{Kurt Gottfried, \AJP\,68 (2000) 143.}

\bibitem{ADY}
{Ady Stern, Yakir Aharonov and Yoseph Imry, Phys Rev. A41 (1990) 3436.}
\bibitem{pwhichfin} {Harry J. Lipkin,
hep-ph/9907551,
Physics Letters  B 477 (2000) 195
and references therein.}  

\bibitem{pnow98}{ Harry J. Lipkin, hep-ph/9901399,
in Proceedings of the Europhysics Neutrino
Oscillation Workshop (NOW'98) 7-9 September 1998. Amsterdam. Published in
http://www.nikhef.nl/pub/conferences/now98/.}

\bibitem{Leo}{Leo Stodolsky, Phys. Rev. D58 (1998) 036006.}



\bibitem{Kayser}{B. Kayser, \PRD\,24 (1981) 110.}




\end{thebibliography}
\end{document}